\documentclass[prl,aps,twoside,showpacs,superscriptaddress,twocolumn,floatfix,reprint,pdflatex]{revtex4}

\usepackage{color}

\usepackage{latexsym}
\usepackage{graphicx}
\usepackage{times}
\usepackage{bbold}

\newcommand{\arccot}{\mathop{\mathrm{arccot}}}

\begin{document}

\title{Experimental implementation of the optimal linear-optical controlled phase gate}

\author{K.\ Lemr}

\affiliation{Joint Laboratory of Optics of Palack\'{y} University and
     Institute of Physics of Academy of Sciences of the Czech Republic,
     779\,07 Olomouc, Czech Republic}

\author{A.\ \v{C}ernoch}

\affiliation{Joint Laboratory of Optics of Palack\'{y} University and
     Institute of Physics of Academy of Sciences of the Czech Republic,
   779\,07 Olomouc, Czech Republic}
 
\author{J.\ Soubusta}

\affiliation{Joint Laboratory of Optics of Palack\'{y} University and
     Institute of Physics of Academy of Sciences of the Czech Republic,
   779\,07 Olomouc, Czech Republic}

\author{K.\ Kieling}
\affiliation{Institute of Physics and Astronomy, University of Potsdam,
     14476 Potsdam, Germany}
 
\author{J.\ Eisert}

\affiliation{Institute of Physics and Astronomy, University of Potsdam,
     14476 Potsdam, Germany}
\affiliation{Institute for Advanced Study Berlin, 14193 Berlin, Germany}
 
\author{M.\ Du\v{s}ek}

\affiliation{Department of Optics, Faculty of Science, Palack\'y University,
   771\,46 Olomouc, Czech~Republic}

\date{\today}

\begin{abstract}
We report on the first experimental realization of optimal
linear-optical controlled phase gates for arbitrary phases. The
realized scheme is entirely flexible in that the phase shift can be tuned
to any given value. All such controlled phase gates are optimal in
the sense that they operate at the maximum possible success
probabilities that are achievable within the framework of any postselected
linear-optical implementation.  The quantum gate is implemented 
using bulk optical elements and polarization encoding of qubit states.
We have experimentally explored
the remarkable observation that the optimum success probability is
not monotone in the phase.
\end{abstract}


\maketitle


Linear-optical architectures belong to the most prominent
platforms for realizing protocols of quantum information
processing \cite{kni01,Further}. In small-scale applications of
quantum information, such as in quantum repeaters, they
will quite certainly play a key role. Unsurprisingly, a
significant research effort has been dedicated in recent years to
experimental realization of universal linear-optical quantum
gates.
Linear-optical quantum gates are probabilistic by their very nature
\cite{kni01}. Therefore, the exact trade-offs between
properties of a gate, such as entangling power, and its
probability of success are in the focus of attention. 

In this work, we explore this
trade-off for the first time experimentally.
We present data from an experimental realization of a linear-optical, 
post-selected controlled phase gate implementing 
the following operation on two qubits:
\begin{equation}
\begin{array}{lcl}
 |0, 0 \rangle &\mapsto& u_{0,0} |0,0\rangle = |0, 0 \rangle , \\
 |0, 1 \rangle &\mapsto& u_{0,1} |0,1\rangle = |0, 1 \rangle , \\
 |1, 0 \rangle &\mapsto& u_{1,0} |1,0\rangle = |1, 0 \rangle , \\
 |1, 1 \rangle &\mapsto& u_{1,1} |1,1\rangle = e^{i \varphi}|1, 1 \rangle,
\end{array}
\label{eqnn:gate}
\end{equation}
for an arbitrary given phase $\varphi\in[0,\pi]$. 
It is key to this experiment that this angle can be 
chosen in a fully tunable fashion, hence adding a 
flexible scheme to the linear optical quantum-information-processing  
toolbox.

Controlled phase gates are important members of this
toolbox. For example, they play a key
role in the circuit for quantum Fourier transform \cite{nie00}.
They are entangling quantum gates in general and,
together with single-qubit operations, they form a universal set for
quantum computing. Notice that the controlled-NOT
gate can be obtained by applying a Hadamard transform to the
target qubit before and after a controlled phase gate with
phase shift $\pi$.

Previous experimental work was devoted to the linear-optical
realization of a special case of the controlled phase gate with
the fixed phase $\varphi=\pi$ \cite{hof02}. Ref.~\cite{lan09} presents an
experiment with phases different from $\pi$, but with a
non-optimal probability of success. The optimal success
probability has recently been identified theoretically in Ref.\ \cite{kie10}. 
This optimum probability
we have indeed reached in the experiment
described in this Letter. We observe the quite remarkable
trade-off between the entangling power of the gate and its success probability, which
is---surprisingly---not monotonous in the phase on
$[0,\pi]$. The success probability decreases rapidly for small
phases, but remains almost constant for phases between
$\pi/4$ and $\pi$. This experiment is hence expected to be both interesting conceptually 
as well as technologically, since a fully tunable bulk linear-optical architecture is presented, 
uplifting tunable schemes for quantum state preparation \cite{Weinfurter}
to the level of quantum information processing.

\begin{figure}
  \begin{center}
  \resizebox{\hsize}{!}{\includegraphics*{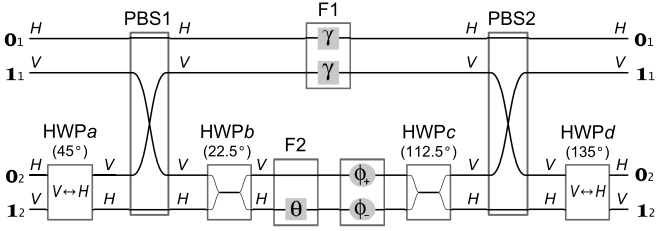}}
  \end{center}
  \caption{Conceptual scheme of the gate. Vertically ($V$)
  and horizontally ($H$) polarized components of the same beam
  are drawn separately for clarity. In polarization beam splitters
  PBS1 and PBS2 the vertical components are reflected. Half-wave
  plates HWP$b$ and HWP$c$ act as ``beam splitters'' for $V$ and $H$
  polarization modes. F1 and F2 are filters (attenuators), F1 acts on the both polarization
modes, F2 on the H component only.
  Phase shifts $\phi_{+}$ and $\phi_{-}$ are introduced by proper
  path differences in the respective modes.
  HWP$a$ and HWP$d$ just swap vertical and horizontal
  polarizations. In the final setup they are omitted for simplicity
  and the second qubit is encoded inversely with respect to the first qubit.}
  \label{scheme1}
\end{figure}

{\it Theoretical framework:} We start by introducing the theoretical framework underlying the 
experiment.
For post-selected linear optical gates, the beam splitter
matrix $A$ describing a general linear optics
network is constrained by
the action of the gate~(\ref{eqnn:gate}) as
\begin{eqnarray*}
	\text{per}A[c_i,c_j|c_k,c_l]=u_{i,j}\delta_{i,k}\delta_{j,l}, 
\end{eqnarray*}
where $i,j,k,l=0,1$.
The left hand side is the permanent of a matrix filled by matrix
elements of $A$~\cite{Scheel04}. $c_0=(0,1)$, $c_1=(1,0)$ are
vectors describing the usual dual-rail encoding into Fock states
on two modes~\cite{Kieling08}. Due to post-selection,
only outcomes in the computational subspace $\mathrm{span}\{|c_k,c_l\rangle\}$ are
considered, giving rise to $16$ quadratic equations in the matrix
elements of $A$. The solutions are (up to mode-permutations) all
of the form $A = \mathbb 1_2\oplus B$ with $B$ being a
$2\times2$-matrix describing the interaction of the two logical-$1$ modes
while the others are just by-passed. Using the singular value
decomposition, this matrix can be expressed as
$B=V\text{diag}(\sigma_+,\sigma_-) W$, where the unitaries $V$ and $W$ have
immediate physical meaning: they describe beam splitters between
the two modes. After rescaling the diagonal matrix by the largest singular value $\sigma_+$, an interpretation of $\text{diag}(1,\sigma_-/\sigma_+)$ can 
be given as well: One
mode is left unaltered while the second one is damped by
$\theta=\sigma_-/\sigma_+$. The success probability is then given by
$p_{\text s} = \sigma_+^{-4}$. 
It turns out \cite{kie10} that for $\varphi \in [0,\pi]$ this
optimal success probability  takes the
form
\begin{eqnarray}\nonumber
   p_{\text s}(\varphi) =
   \left(1+2\left|\sin\frac{\varphi}{2}\right|+2^{3/2}\sin\frac{\pi-\varphi}{4}{\left|\sin\frac{\varphi}{2}\right|^{1/2}}\right)^{-2}
\end{eqnarray}
and $W=\left[\begin{array}{cc} -1&-1 \\ 1&-1
\end{array}\right]/\sqrt{2}$, $V=W^{-1}\text{diag}(\mathrm
e^{\imath\phi_+},\mathrm e^{\imath\phi_-})$, where the phases in
the lower Mach-Zehnder interferometer (between HWP$b$ and HWP$c$, see
Fig.~\ref{scheme1}) are defined by
\begin{eqnarray*}
  \phi_{\pm} = \arccot \left[ \cot\frac{\varphi+\pi}{4} \pm
               \left( (2 - 2\cos \varphi)^{1/4} \sin\frac{\varphi+\pi}{4}
               \right)^{-1} \right].
\end{eqnarray*}
Phase shifts are applied to each arm
and one arm is damped by an attenuator (neutral-density filter) which
is characterized by an amplitude transmissivity of
\begin{eqnarray*}
  \theta = \left({\frac{1 + 2\sin\frac{\varphi}{2} - 2(2 - 2\cos \varphi)^{1/4}
                          \cos\frac{\varphi+\pi}{4}}
                      {1 + 2\sin\frac{\varphi}{2} + 2(2 - 2\cos \varphi)^{1/4}
                          \cos\frac{\varphi+\pi}{4}}}\right)^{1/2}.
\end{eqnarray*}
The remaining two attenuators in the
upper beams are used to damp the amplitude of the by-passed modes
to compensate for the overall losses in the lower beams. Their amplitude transmissivity reads $\gamma=p_{\text
s}^{1/4}$.


{\it Details of the experiment:}
As the starting point of this experiment we generate a pair of photons
in the process of type-I spontaneous parametric down-conversion.
The laser beam of 250\,mW of cw optical power emitted by
Krypton-ion laser at 413\,nm impinges on the LiIO$_3$ crystal. Pairs
of photons at 826\,nm are collected using single mode fibres
serving also as spatial filters. Subsequently, polarization
controllers are employed to adjust the horizontal polarization of
the photons.


\begin{figure}
  \begin{center}
  \resizebox{\hsize}{!}{\includegraphics*{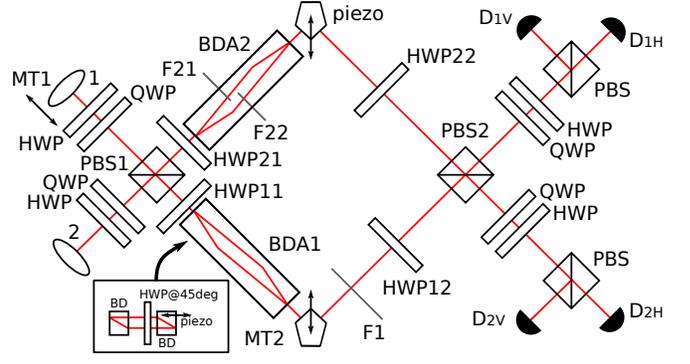}}
  \end{center}
  \caption{Scheme of the actual experimental setup (see text for details).}
  \label{setup}
\end{figure}


The half-wave plates (HWP) and quarter-wave plates (QWP) in the
input arms (see Fig.~\ref{setup}) are used to set the input
states. Subsequently, the photons are superposed on the first
polarizing beam splitter PBS1 which transmits horizontal and
reflects vertical polarization. Due to imperfections the
transmissivity for horizontal polarization is only $95$\,\% (the
remaining $5$\,\% are reflected). Polarization beam splitters also
introduce parasitic phase shifts between vertical and horizontal
polarization components. After leaving the PBS1 the photons in the
upper arm are subjected to the action of half-wave plate HWP21.
When set to $22.5$\,deg it performs the
transformation $|H \rangle \mapsto  ( |H\rangle + |V\rangle
)/\sqrt{2}$, $|V\rangle \mapsto (| H \rangle- |V \rangle )/\sqrt{2}$,
where $|H\rangle$ and $|V\rangle$ denote horizontal and vertical polarization states,
respectively. The lower arm is also equipped with a half
wave-plate (HWP11) but it is set to zero (its
presence just guarantees the same optical paths, dispersion
effects, etc.\ in the both arms). Behind the wave plates there are
the beam-divider assemblies BDA1 and BDA2. They consist of two
beam dividers (BD) splitting and subsequently rejoining horizontal and
vertical polarizations. BDA2 is equipped with gradient
neutral-density filters F21 and F22 (see Fig.~\ref{setup}). This
way one can perform arbitrary polarization sensitive losses. BDA1
is used just to equilibrate the beam position and the optical
length of the both arms. After leaving the beam-divider assemblies
the photons propagate through half-wave plates HWP12 and HWP22.
HWP22 is set to $22.5$\,deg reversing thus the transformation
imposed by HWP21. HWP12 is set to $45$\,deg to compensate for the
polarization flip between the $H$ and $V$ polarizations performed
by BDA1. The lower arm is equipped with a gradient neutral density filter F1 to
apply polarization independent losses. The
gate operation itself is completed by overlapping the photons on
the second polarizing beam splitter PBS2. To be able to perform
complete state and thereby {\it process tomography} we employ polarization
analysis in the both output arms. The analysis consists of QWPs
and HWPs followed by polarizing beam splitters, cut-off filters
and single mode fibres leading to single photon detectors.

{\it Gate operation:} 
The setup is then adjusted to perform the gate operation. First we set filters
F21 and F1 to introduce the required losses.
After that the wave plates HWP21 and HWP22 are
set to $22.5$\,deg. The phase in the beam divider assembly BDA2 is
set to maximize the visibility of the interferometer formed by
PBS1 and PBS2.
The precise tuning of the gate is then performed by switching
between the inputs $|H_1, R_2\rangle$ and $|V_1 ,R_2\rangle$,
where indices $1$ and $2$ denote the input modes and $R$ stands for the
right circular polarization. Using the circular detection basis in
the second output arm we can observe the phase applied
by the gate when the polarization of the first input photon flips from
$|H\rangle$ to $|V\rangle $. In this configuration we also tune the phase shift inside
the beam divider assembly BDA2 and the phase shift between the two
arms of the Mach-Zehnder interferometer formed by PBS1 and PBS2.

\begin{figure}
  \begin{center}
  \resizebox{\hsize}{!}{\includegraphics*{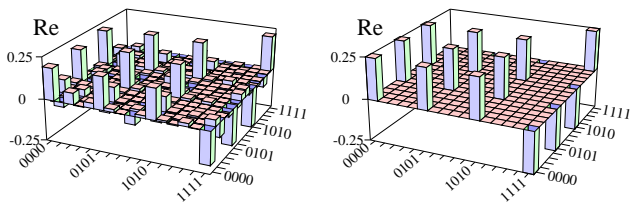}}
  \end{center}
  \caption{Matrix representation of the CP map characterizing the operation of the
           controlled-phase gate with $\varphi=\pi$. The left panel shows
           the real part of the reconstructed process matrix, the right one
           displays the real part of the ideal theoretical CP map.
           Imaginary parts are negligible (zero in ideal case). The process fidelity $F_{\chi}=84$\%.}
  \label{mat1}
\end{figure}

\begin{figure}
  \begin{center}
  \resizebox{\hsize}{!}{\includegraphics*{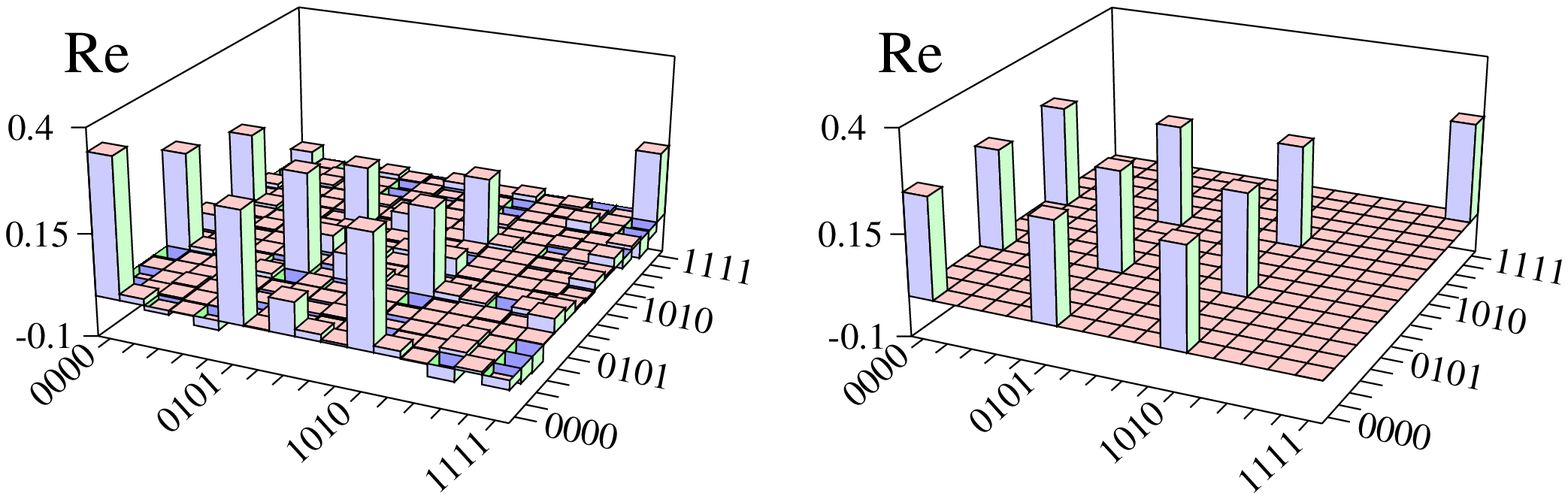}}
  \resizebox{\hsize}{!}{\includegraphics*{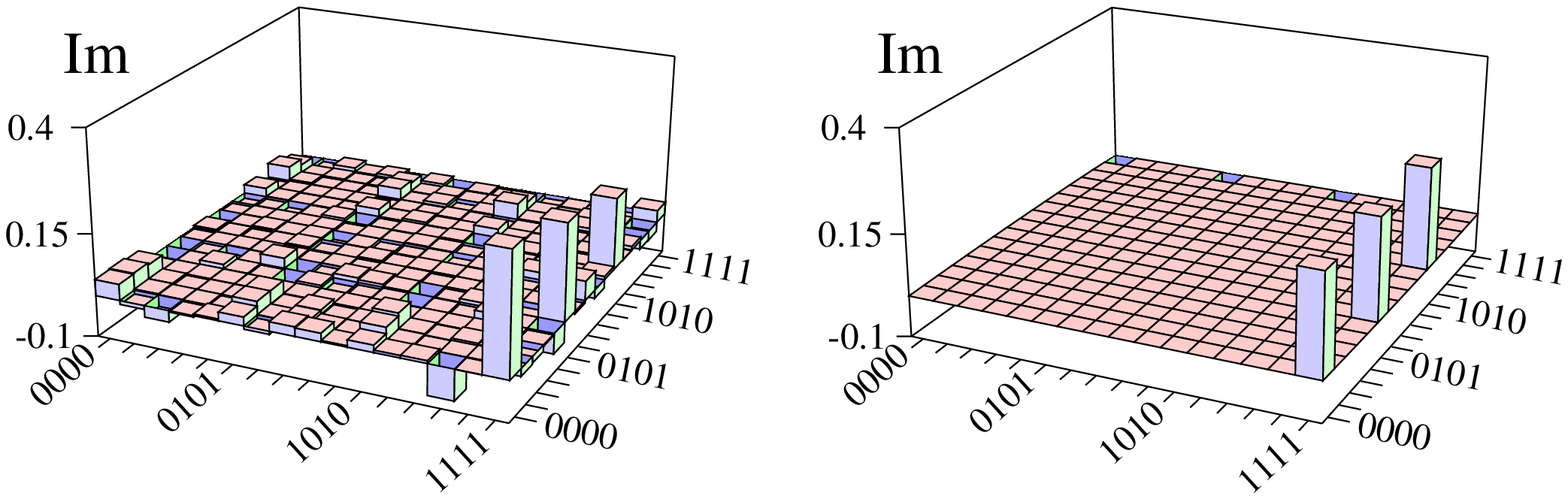}}
  \end{center}
  \caption{Choi matrices for the gate with $\varphi=\pi/2$. The left top panel shows
           the real part of the reconstructed process matrix while the left bottom one
           displays its imaginary part. The process fidelity $F_{\chi}=86$\%. The two right panels show the ideal matrix.}
  \label{mat2}
\end{figure}

{\it Results:}
Gradually we have adjusted the gate to apply $7$ phases in the range between $0$ and $\pi$.
Each time we have performed complete process tomography and
estimated the process matrix using the maximum likelihood method.
Fidelities of the process lie in the range from $84$\% to $95$\% (see
Tab.~\ref{tab1}). Figs.~\ref{mat1} and \ref{mat2} show
examples of experimentally obtained process matrices and their
theoretical counterparts for $\varphi=\pi$ and $\pi/2$.


\begin{table}
  \begin{ruledtabular}
  \begin{tabular}{cccccccc}
  $\varphi$ & $F_{\chi}$ & $F_{\mathrm{av}}$ & $F_{\mathrm{min}}$ &
  $\mathcal{P}_{\mathrm{av}}$ & $\mathcal{P}_{\mathrm{min}}$ &
  $p_{\mathrm{s,obs}}$ & $p_{\mathrm{s,th}}$
  \\ \hline
  0          & 0.94 & 0.96 & 0.84 & 0.96 & 0.87 & 0.87 $\pm$ 0.08 & 1.00 \\
  $0.05\pi$  & 0.95 & 0.96 & 0.91 & 0.96 & 0.87 & 0.37 $\pm$ 0.05 & 0.36 \\
  $0.125\pi$ & 0.91 & 0.90 & 0.77 & 0.95 & 0.87 & 0.19 $\pm$ 0.03 & 0.21 \\
  $0.25\pi$  & 0.84 & 0.88 & 0.73 & 0.90 & 0.67 & 0.11 $\pm$ 0.02 & 0.13 \\
  $0.5\pi$   & 0.86 & 0.89 & 0.81 & 0.90 & 0.76 & 0.09 $\pm$ 0.01 & 0.09 \\
  $0.75\pi$  & 0.84 & 0.87 & 0.63 & 0.90 & 0.71 & 0.08 $\pm$ 0.01 & 0.09 \\
  $\pi$      & 0.84 & 0.86 & 0.71 & 0.92 & 0.83 & 0.12 $\pm$ 0.01 & 0.11 \\
\end{tabular}
\end{ruledtabular}
  \caption{Process fidelities ($F_{\chi}$), average ($F_{\mathrm{av}}$)
           and minimal ($F_{\mathrm{min}}$) output-state fidelities,
           average ($\mathcal{P}_{\mathrm{av}}$) and minimal ($\mathcal{P}_{\mathrm{min}}$)
           output-state purities and actually observed ($p_{\mathrm{s,obs}}$)
           and theoretically predicted ($p_{\mathrm{s,th}}$) success probabilities for
           different phases ($\varphi$).}
           \label{tab1}
\end{table}


For each selected phase we simultaneously measured
two-photon coincidence counts between detectors $\rm D_{1H} \&
D_{2H}$, $\rm D_{1V} \& D_{2V}$, $\rm D_{1H} \& D_{2V}$, and $\rm
D_{1V} \& D_{2H}$, each for $3 \times 3$ combinations of
polarization measurement bases in the output arms. This amounts to
measuring projections onto horizontal/vertical,
diagonal/anti-diagonal and right/left circular
polarizations. The diagonal and anti-diagonal linear polarization
states are defined as
$|D\rangle= (|H\rangle+|V\rangle)/\sqrt{2}$ and
$|A\rangle= (|H\rangle-|V\rangle)/\sqrt{2}$ and the right-
and left-handed circular polarization states read
$|R\rangle= (|H\rangle+i|V\rangle)/\sqrt{2}$,
$|L\rangle= (|H\rangle-i|V\rangle)/\sqrt{2}$. The unequal
detector efficiencies were compensated by proper re-scaling of the
measured coincidence counts \cite{Soubusta07}. Each measurement
was done for $36$ different input product states. Namely, for $6
\times 6$ combinations of polarization state vectors $|H\rangle$,
$|V\rangle$, $|D\rangle$, $|A\rangle$, $|R\rangle$, and
$|L\rangle$ of each input photon. This complex measurement
provided us with tomographically complete data enabling us to fully
characterize the implemented operation by quantum process
tomography \cite{Poyatos97,Jezek03}
as well as to reconstruct density matrices of output states for
each used input state.

{\it Active stabilization:} Each setting of an input and output polarization basis was preceded by an active stabilization. For the purpose of the stabilization the fixed input state and output detection basis were always used. In this setting the visibility in the interferometer formed by PBS1 and PBS2 was measured. If this visibility was lower than a selected threshold then the positions of MT1 (interferometer lengths) and MT2 (dip position) were optimized and the phase drift was compensated. Finally the required polarizations were set and data were accumulated within $5$\,s.

{\it Process tomography:}
Any quantum operation can be fully described by a completely
positive map and---according to the Jamiolkowski-Choi isomorphism---represented
by a positive-semidefinite operator
$\chi$ on the tensor product of input and output Hilbert spaces
\cite{jam72}. In our case $\chi$ is a $16\times16$ square matrix.
From
the measured data we can reconstruct $\chi$ for any setting of
$\varphi$ using maximum likelihood estimation
\cite{Jezek03,Paris04}. To quantify the quality of the operation we calculate the process
fidelity, if $\chi_{\mathrm{id}}$ is a one-dimensional projector its
common definition is 
$F_\chi=\mathrm{Tr}[\chi \chi_{\mathrm{id}}]/(\mathrm{Tr}[\chi]\mathrm{Tr}[\chi_{\mathrm{id}}])$.
Here $\chi_{\mathrm{id}}$ represents the ideal transformation
corresponding to the controlled-phase gate. Specifically,
\begin{eqnarray*}
\chi_{\mathrm{id}} = \sum_{i,j,k,l=V,H} |i,j \rangle\langle k,l|
\otimes U | i,j \rangle\langle k,l | U^\dag, 
\end{eqnarray*}
where $U$ stands for
the unitary operator on two qubits defined by
Eq.~(\ref{eqnn:gate}).
We have also reconstructed the density matrices of output
two-photon states corresponding to all product inputs
$|j, k\rangle$, $j,k\in\{H,V,D,A,R,L\}$.
This was done for all values of $\varphi$.
An important parameter characterizing the gate performance is the
fidelity of output states $\rho_{\mathrm{out}}$ defined as
$F=\langle \psi_{\mathrm{out}}| \rho_{\mathrm{out}}
|\psi_{\mathrm{out}} \rangle$, where $|\psi_{\mathrm{out}} \rangle
= U |\psi_{\mathrm{in}} \rangle$ and $|\psi_{\mathrm{in}} \rangle$
is the input state vector. Table~\ref{tab1} contains the average
and minimal values of state fidelities for different phases.
Fidelities $F_{\mathrm{av}}$ are averaged over all output states
corresponding to our $36$ input states; $F_{\mathrm{min}}$ denote minimal values.
Another important characteristics is the purity of the output
state $\rho_{\mathrm{out}}$, defined as
$\mathcal{P}=\mathrm{Tr}[\rho_{\mathrm{out}}^2]$. If the input
state is pure the output state is expected to be pure as well. The
average and minimal purities of output states are also given in
Table~\ref{tab1}.

\begin{figure}[t]
  \begin{center}
  \resizebox{\hsize}{!}{\includegraphics*{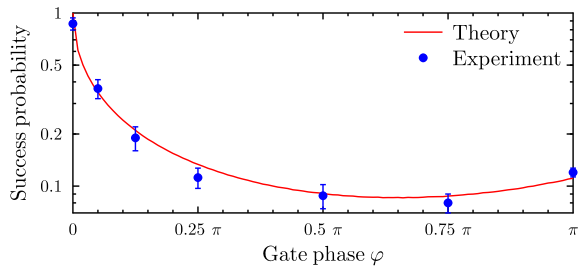}}
  \end{center}
  \caption{Success probability of the gate. 
           	}
  \label{fig-p_s}
\end{figure}

{\it Trade off in success probabilities:} The most important result of this paper---aside from
the technological implications---is the experimental
verification of the trade-off between the phase shift applied by
the gate and the corresponding success probability of the gate.
We have estimated the success probability
for each value of the selected phase shifts.
It was calculated as a
ratio of the number of successful gate operations per time
interval and the number of reference counts during the same
interval (measured with no filters and with the wave plates set to 0).
We have determined the success probability for the all selected input states.
These probabilities were averaged and the standard deviations were calculated.
Notice that the calibration measurements collect coincidence
counts behind the setup (using the same detectors as in the
subsequent measurements), thus all the ``technological`` losses in the setup and low
detector efficiencies are included in the calibration.
Therefore the estimated success probabilities are not
burdened by these ``technological'' losses. They
can be compared with the
theoretical predictions in Tab.~\ref{tab1} and in Fig.~\ref{fig-p_s}.
One can see a very good agreement with the theoretical prediction.


\emph{Conclusions:\/} We have built the first implementation of the tunable
linear-optical controlled phase gate which is optimal for any
value of the phase shift. Changing the parameters of the setup the
gate can apply any phase shift from the interval $[0,\pi]$ on the
controlled qubit. We have thoroughly tested the performance of the
gate using full quantum process tomography. Obtained
process fidelities range from $84$\% to $95$\%.
We have also experimentally verified that all our controlled phase
gates are optimal in the sense that they operate at the maximum
possible success probabilities that are achievable by
linear-optical setups. The experimental verification of this
trade-off between the phase shift applied by the gate and the
corresponding success probability of the gate is the most notable
result of our work. It demonstrates the contra-intuitive fact that the
optimal success probability is not monotonous with the phase shift
increasing from 0 to $\pi$. It is the hope that the flexible tool established here
proves useful in devising further optical linear optical circuits for optical
quantum information processing and that ideas
developed in this work find their way to realization in fully integrated optical
architectures.

{\it Acknowledgments:}
This work was supported by the Czech Ministry of Education (1M06002, MSM6198959213), the Czech Science Foundation (202/09/0747), the EU (QESSENCE, MINOS, COMPAS), the
EURYI and the Palacky University (PrF-2010-009 and PrF-2010-020).

\end{document}